\input epsf

\documentstyle[prd,floats,aps]{revtex}

\tighten

\begin{document}

\draft

\twocolumn[\hsize\textwidth\columnwidth\hsize\csname @twocolumnfalse\endcsname
\preprint{AEI-1999-7}

\title{Nonlinear and Perturbative Evolution of Distorted Black Holes.  II.  
Odd-parity Modes}

\author{John Baker$^{(1)}$, Steven Brandt$^{(4)}$, Manuela Campanelli$^{(1)}$,
 Carlos O. Lousto$^{(1,5)}$,\\
Edward Seidel$^{(1,2,3)}$ and Ryoji Takahashi$^{(1)}$}

\address{
$^{(1)}$Albert-Einstein-Institut,
Max-Planck-Institut f{\"u}r Gravitationsphysik,
Am M\"uhlenberg 5, D-14476 Golm, Germany\\
$^{(2)}$National Center for Supercomputing Applications,
Beckman Institute, 405 N. Mathews Ave., Urbana, IL 61801 \\
$^{(3)}$Departments of Astronomy and Physics,
University of Illinois, Urbana, IL 61801 \\
$^{(4)}$Department of Astronomy and Astrophysics and Center for
Gravitational Physics and Geometry, \\
Pennsylvania State University, University Park, PA 16802\\
$^{(5)}$Instituto de Astronom\'{\i}a y F\'{\i}sica del Espacio--CONICET,
Buenos Aires, Argentina
    }

\date{\today}

\maketitle

\begin{abstract}
We compare the fully nonlinear and 
perturbative evolution of nonrotating
black holes with 
odd-parity distortions utilizing the perturbative
results to interpret the nonlinear results. 
This introduction of the second polarization (odd-parity) mode of the system, 
and the systematic use of combined techniques brings us 
closer to the goal of studying more complicated systems
like distorted, rotating black holes, such as those formed in the 
final inspiral stage of two black holes.
The nonlinear evolutions are performed with 
the 3D parallel code for Numerical Relativity, {\sl Cactus}, and an 
independent axisymmetric code, {\sl Magor}.  The linearized calculation 
is performed in two ways: (a) We 
treat the system as a metric perturbation on Schwarzschild, using the 
Regge-Wheeler equation to obtain the waveforms produced.  
(b) We treat the system as a curvature perturbation of a Kerr 
black hole (but here restricted to the case of vanishing rotation 
parameter $a$) and evolve it with the Teukolsky 
equation 
The comparisons of the waveforms obtained show an excellent 
agreement in all cases.

\end{abstract}


\vskip2pc]



\section{Introduction}
\label{sec:Introduction}

Coalescing black holes are considered one of the most promising 
sources of gravitational waves for gravitational wave observatories 
like the LIGO/VIRGO/GEO/TAMA network under construction (see, e.g., 
Ref.~\cite{Flanagan97a,Flanagan97b} and references therein).  Reliable 
waveform information about the merger of coalescing black holes can be 
crucial not only to the interpretation of such observations, but also 
could greatly enhance the detection rate.  Therefore, it is crucial to 
have a detailed theoretical understanding of the coalescence process. 

It is generally expected that full scale, 3D numerical relativity will 
be required to provide such detailed information.  However, numerical 
simulations of black holes have proved very difficult.  Even in 
axisymmetry, where coordinate systems are adapted to the geometry of 
the black holes, black hole systems are difficult to evolve beyond 
about $t=150M$, where $M$ is the mass of the system\cite{Anninos94b}.  
In 3D, the huge memory requirements, and instabilities associated 
presumably with the formulations of the equations themselves, make 
these problems even more severe.  The most advanced 3D calculations 
based on traditional Cauchy evolution methods published to date, 
utilizing massively parallel computers, have difficulty evolving 
Schwarzschild~\cite{Anninos94c}, Misner~\cite{Anninos96c}, or 
distorted Schwarzschild~\cite{Camarda97b} beyond about $t=50M$.  
Characteristic evolution methods have been used to evolve distorted 
black holes in 3D indefinitely\cite{Gomez98a}, although it is not 
clear whether the technique will be able to handle highly distorted or 
colliding black holes due to potential trouble with caustics.

In spite of such difficulties, much physics has been learned and 
progress has been made in black hole simulations, in both axisymmetry 
and in 3D. In axisymmetry, calculations of distorted black holes with 
\cite{Brandt94a,Brandt94b,Brandt94c} and without angular 
momentum\cite{Abrahams92a,Bernstein93b}, Misner two black hole initial 
data, including variations of boosted and unequal mass black holes, 
have been all been successfully carried 
out\cite{Anninos94b,Anninos93b,Anninos98a}, and the waveforms 
generated during the collision process have been extensively compared 
to calculations performed using perturbation 
theory\cite{Baker96a,Price94b,Pullin98a,Allen97a}.  In 3D, similar 
calculations have been carried out, especially in evolution of 3D 
distorted black holes~\cite{Allen98a} where it was shown that very 
accurate waveforms can be extracted as a distorted black holes settles 
down, as is expected to happen when two black holes coalesce.  

One of the important results to emerge from these studies is that the 
full scale numerical and perturbative results agree very well in the 
appropriate regimes, giving great confidence in both approaches.  In 
particular, the perturbative approach turned out to work extremely 
well in some regimes where it was not, {\em a priori}, expected to be 
accurate.  For example, in the head-on collision of two black holes 
(using Misner data), the perturbative results for both waveforms and 
energy radiated turned out to be remarkably accurate against full 
numerical simulations -- even in some cases where the black holes had 
distinct apparent horizons. These impressive agreements have 
then been improved by the use of second order perturbation 
theory (see Ref.~\cite{Pullin98a} for a comprehensive review on the 
Zerilli approach and Ref.~\cite{Campanelli99} for the more recent 
curvature based approach holding also for rotating black holes). 
Study of perturbations also offered the plausible explanation that 
the peak of the potential barrier that surrounds a black hole was 
the more relevant quantity, not the horizon.
In a more complex application, the collision of boosted 
black holes was studied.  With a small boost, the total energy 
radiated in the collision was shown to go {\em down} when compared 
with the Misner data.  Linear theory was able to show that there were 
two components to the radiation, one from the background Misner 
geometry and one from the boost.  These two components are 
anti-correlated and combined to produce what has since been called the 
``Baker dip''\cite{Baker96a}.  Had there only been a perturbative 
analysis one might have worried that nonlinear effects might eliminate 
the dip.  Had there been only a full numerical simulation, the dip 
might have been thought to be evidence of a coding error.  When the 
two were combined, however, a confirmation of the correctness of both 
procedures was established and the effect understood.

These are just two examples of a rather large body of work that has 
led to a revival of perturbative calculations, now considered to be 
used as a tool to aid in the verification and interpretation of 
numerically generated results.  The potential uses of this synergistic 
approach to black hole hole evolutions, combining both numerical and 
perturbative evolutions, are many.  First, the two approaches go 
hand-in-hand to verify the full scale nonlinear numerical evolutions, 
which will become more and more difficult as 3D binary mergers of 
unequal mass black holes are attempted, with linear, spin, and orbital 
angular momentum.  Second, as the above examples show, they can aid 
greatly in the interpretation and physical understanding of the 
numerical results, as also shown in 3D distorted black hole 
evolutions~\cite{Allen98a}.  Such insight will become more important 
as we move towards more complex simulations.  (As an example of this 
below, we will show how nonlinear effects and mode-mixing can be 
understood and cleanly separated  from linear effects with this 
approach.) 

Finally, there are at least two important ways in which a 
perturbative treatment can actually aid the numerical simulation.  
First, as shown in Ref.~\cite{Abrahams97a}, it is possible to use 
perturbative evolutions to provide good outer boundary conditions for 
a numerical simulation, since away from the strong field region one 
expects to see low amplitude gravitational waves propagating on a 
black hole background.  This information can be exploited in the 
outer region in providing boundary data.  Second, this combined 
approach can be used in future applications of perturbative approaches 
to ``take over'' and continue a previously computed full scale 
nonlinear numerical simulation.  For example, if gravitational 
waveforms are of primary interest in a simulation, once the system has 
evolved towards a perturbative regime (e.g., two coalescing black 
holes form a distorted Kerr hole, or evolve close enough that a close 
limit approximation is valid), then one may be able to extract the 
relevant gravitational wave data, and evolve them on the appropriate 
black hole background to extract waveforms~\cite{Abrahams95d}.  Not only 
would such a procedure save computational time, it may actually be 
necessary in some cases to extend the simulations.  As discussed 
above, 3D black hole evolutions using traditional ADM style 
formulations, with singularity avoiding slicings, generally break down 
before complete wave forms can be extracted.  A perturbative approach 
may be necessary in such cases to extract the relevant waveform 
physics. This work (called {\it Lazarus project}) is currently being
undertaken by some of the present authors.

However, all work to date in this area of comparing full scale 
numerical simulations with perturbative approaches has dealt with 
even-parity distortions of Schwarzschild-like black holes.
See for instance Ref.~\cite{Allen97a}, referred to here as Paper I, 
where we compared perturbative techniques, based on the Zerilli 
approach, with fully nonlinear evolutions of even-parity distorted 
black holes. This restriction to the Zerilli equation cannot handle 
the odd-parity class of perturbations, and more importantly, 
it cannot be applied easily to the case of rotating black holes.  
The more general black hole case has both even- 
and {\em odd-}parity distortions, and also involves black holes with 
angular momentum. For this reason, in this paper we take 
an important step towards application to the more general case 
of rotating, distorted black holes, by introducing the Teukolsky 
equation as the fundamental perturbation equation. 
In fact, for black holes with angular momentum, 
there is not an $\ell -m$ multipole decomposition of metric 
perturbations in the time domain and the most natural way to proceed 
is with the curvature-based perturbation formalism leading to the 
Teukolsky equation, which also simultaneously handles, in a completely
gauge invariant way, both even- and odd-parity perturbations.

The paper is structured as follows. In section II, we review the 
initial data sets and {\it four} different techniques and 
approaches to evolve black holes. 

\begin{enumerate}

\item We first carry forward the metric-based perturbation approach 
by considering the Regge-Wheeler (odd-parity) equation to perform 
perturbative evolutions, and for the first time apply these techniques 
to a class of distorted black hole data sets containing even-and 
odd-parity distortions.  

\item We also show how one can carry out such perturbative 
evolutions with the curvature-based Teukolsky equation, using the 
same initial datasets. Although in certain cases, the metric perturbations 
can be computed from the curvature perturbations, and vice versa
~\cite{Campanelli98b}, in general using both approaches  
helps us better to understand the systems we are dealing with.

\item We carry out fully nonlinear evolutions of the same data sets for 
comparison with a 2D (axisymmetric) code, {\sl Magor}, also 
capable of evolving distorted rotating black holes. 

\item Finally, the same initial data is evolved in its full 3D mode 
with a general parallel code for numerical relativity, {\sl Cactus}.

\end{enumerate}

In section III we finally discuss the results in detail 
and show how the combination of these different approaches 
provides an extremely good and systematic strategy to 
cross check and further verify the accuracy of the codes used.
The comparisons of waveforms obtained in this way show an 
excellent agreement, in both perturbative and full nonlinear 
regimes.   

Although in this paper we restrict ourselves to the case of initial 
datasets without angular momentum, the family of datasets we 
use for this study also includes distorted Kerr black holes, which 
will be considered in a follow up paper. In fact, our eventual goal 
is to apply both fully nonlinear numerical and perturbative techniques 
to evolve a binary black hole system near the merger phase, which 
final stage can be reasonable modeled by a single distorted Kerr 
black hole. In this case we should be able to address extremely 
important questions like how much energy and angular momentum 
can be radiated in the final merger stage of two black holes.

\section{Four Ways to Evolve Distorted Black Holes}

\subsection{Distorted Black Hole Initial Data}
\label{sec:initial}

Our starting point is represented by a distorted black hole initial 
data sets developed originally by Brandt 
and Seidel~\cite{Brandt94a,Brandt94b,Brandt94c} to mimic the 
coalescence process.  These data sets correspond to ``arbitrarily'' 
distorted rotating {\em single} black holes, such as those that will 
be formed in the coalescence of two black holes.  Although this black 
hole family can include rotation, in this first step we restrict 
ourselves to the non-rotating limit (the so-called ``Odd-Parity 
Distorted Schwarzschild'' of 
Ref.~\cite{Brandt94a,Brandt94b,Brandt94c}).  However, these data sets 
do include both degrees of gravitational wave freedom, including the 
``rotation-like'' odd-parity modes.

The details of this initial data procedure are covered 
in~\cite{Brandt94a,Brandt94b,Brandt94c}, so we will go over them only 
briefly here.  We follow the standard 3+1 ADM decomposition of the 
Einstein equations which give us a spatial metric, an extrinsic 
curvature, a lapse and a shift.  We choose our system such that we 
have a conformally flat three--metric $\gamma_{ij}$ defined by
\begin{equation}
    ds^{2}= \Psi^{4} \left(d\eta^{2} + d\theta^{2}+ \sin^{2}\theta 
    d\varphi^{2}\right).
\end{equation}
where the coordinates $\theta$ and $\varphi$ are the usual 
spherical coordinates and the radial coordinate has been replaced by
an exponential radial coordinate$\eta$ 
($\bar{r}=\frac{M}{2}e^\eta$). Thus, if we let the conformal factor
be $\Psi=\sqrt{\bar{r}}$ we have the
flat space metric with the origin at $\eta = -\infty$.  If we let
\begin{equation}
\Psi=\sqrt{\bar{r}}\left(1+\frac{M}{2\bar{r}}\right)
\end{equation}
we have the Schwarzschild 3-metric.  In this case one 
finds that $\eta=0$ corresponds to the throat of a Schwarzschild 
wormhole, $\eta=\pm\infty$ corresponds to spatial infinity in each of 
the two spaces connected by the Einstein-Rosen bridge (wormhole).  
Note also that this metric is invariant under the isometry operation 
$\eta\rightarrow -\eta$.  In the full nonlinear 2D evolution 
will use this fact to give ourselves the appropriate boundary 
conditions for making distorted black holes.

The extrinsic curvature is chosen to be
\begin{eqnarray}
K_{ij} & = & \Psi^{-2} h_{ij} = \Psi^{-2} \left( \begin{array}{ccc}
0 & 0 & H_E \\
0 & 0 & H_F \\
H_E & H_F & 0
\end{array} \right)
\end{eqnarray}
where
\begin{eqnarray}
H_E & = & q_G \left(\left(n^\prime+1\right)-
\left(2+n^\prime \right) \sin^2\theta \right) \sin^{n^\prime-1}\theta \\
H_F & = & -\partial_\eta q_G
\cos\theta \sin^{n^\prime} \theta \\
q_G &=& Q_0\left[\exp\left(-\left(\eta-\eta_0\right)^2/\sigma^2\right)\right.
\nonumber\\
&&\left.
+\exp\left(-\left(\eta+\eta_0\right)^2/\sigma^2\right)\right].
\end{eqnarray}
The various functions have been chosen so that the momentum constraints 
are automatically satisfied, and have the form of odd-parity 
distortions in the black hole extrinsic curvature.  The function 
$q_{G}$ provides an adjustable distortion function, which satisfies 
the isometry operation, and whose amplitude is controlled by the 
parameter $Q_{0}$.  This parameter carries units of length squared.
Since we will be comparing cases with different masses we will refer
to an amplitude $\tilde Q_0=Q_0/M^2$ normalized by the ADM mass of
the initial slice.   If $Q_{0}$ vanishes, an unperturbed Schwarzschild 
black hole results.  The parameter $n^\prime$ is used to describe an 
``odd-parity'' distortion.  It must be odd, and have a value of at 
least 3.  The function $\Psi$ is the conformal factor, which we have 
abstracted from the metric and extrinsic curvature according to the 
factorization given by Lichnerowicz~\cite{Lichnerowicz44}.  This 
decomposition is valuable, because it allows us to solve the momentum 
and Hamiltonian constraints separately (with this factorization the 
extrinsic curvature given above analytically solves the momentum 
constraints).  

For the class of data considered here the only nontrivial component
of the momentum constraints is the $\varphi$ component:
\begin{equation}
\partial_\eta \hat{H}_{E} \sin^3\theta+
\partial_\theta \left( \hat{H}_F \sin^2\theta \right) =0.
\label{mom constraint}
\end{equation}
Note that this equation is independent of the function $q_{G}$.
This enables us to choose the solutions to these equations independently
of our choice of metric perturbation.

At this stage we solve the Hamiltonian constraint numerically to 
obtain the appropriate value for $\Psi$.  Data at the inner boundary 
($\eta=0$) is provided by an isometry condition, namely, that the 
metric should not be changed by an inversion through the throat 
described by $\eta \rightarrow -\eta$.  If we allow $Q_0$ to be zero, 
we recover the Schwarzschild solution for $\Psi$.  

The Hamiltonian constraint equation can be expanded in coordinate form 
to yield, in this case,
\begin{eqnarray}
\frac{\partial^2 \Psi}{\partial\eta^2}&+&\frac{\partial^2 
\Psi}
{\partial\theta^2}+ 
\frac{\partial \Psi}{\partial\theta}\cot\theta -\frac{\Psi}{4} = \nonumber\\
&-&\frac{\Psi^{-7}}{4}\left( \hat{H}_E^2 \sin^2\theta
+\hat{H}_F^2 \right).
 \label{Energy Constraint}
\end{eqnarray}

This construction is similar to that given in Bowen and 
York~\cite{Bowen80}, except that form of the extrinsic curvature is 
different.  The same procedure described above can also be used to 
construct Kerr and distorted Kerr black holes, as described in 
Ref.~\cite{Brandt94a,Brandt94b,Brandt94c}, but we defer that 
application to a future paper.

Note that although the form of the extrinsic curvature is decidedly 
odd-parity (consider reversing the the $\varphi$-direction),
the Hamiltonian constraint equation for $\Psi$ affects the
diagonal elements of the three-metric producing a nonlinear
even-parity distortion.
  If both $\hat{H}_F$ 
and $\hat{H}_E$ vanish, undistorted Schwarzschild results.  If they 
are present, they generate a linear odd-parity perturbation directly 
through the extrinsic curvature, and a second order even-parity 
perturbation through the conformal factor $\Psi$.  Hence, the system 
will have mixed  odd- and even-parity distortions mixed together at 
different perturbative orders.  As we will see below,
 because even- and odd-parity components are cleanly separated 
in this way, and the background 
geometry is explicitly Schwarzschild, it is straightforward to construct 
analytic, linearized initial data for these distorted black holes, which 
can then be evolved with the perturbation equations.

In summary, our initial data sets contain both even- and odd-parity 
distortions of a Schwarzschild black hole, and are characterized by 
parameters $(Q_{0},n',\eta_{0},\sigma)$, where $Q_{0}$ determines the 
amplitude of the distortion,$n'$ determines the angular pattern, 
$\eta_{0}$ determines the radial location (with $\eta_{0}=0$ being the 
black hole throat), and $\sigma$ determines the radial extent of the 
distortion.  For simplicity of discussion, all cases we will consider in 
this paper have the form $(Q_{0},n',\eta_{0}=2,\sigma=1)$.

\subsection{Two Perturbation Formalisms}

\subsubsection{Metric Perturbations}

The theory of metric perturbations around a Schwarzschild hole was 
originally derived by Regge and Wheeler\cite{Regge57} for odd-parity 
perturbations and by Zerilli\cite{Zerilli70a} for even-parity ones.  
The spherically symmetric background allows for a multipole 
decomposition even in the time domain.  Moncrief\cite{Moncrief74} has 
given a gauge-invariant formulation of the problem, which like the 
work of Regge-Wheeler and Zerilli, is given in terms of the 
three-geometry metric perturbations. We will use the Moncrief 
formalism here as already described in Paper I.

For special combinations of the perturbation equations, a 
wave equation, the famous Regge-Wheeler equation, 
resulted for a single function $\phi_{(\ell m)}$:

\begin{equation}\label{rtrw}
-\frac{\partial^2\phi_{(\ell m)}}{\partial t^2}
+\frac{\partial^2\phi_{(\ell m)}}{\partial r*^2}-
V_{\ell}^-(r)\phi_{(\ell m)}=0 \; .
\end{equation}
Here $r^*\equiv r+2M\ln(r/2M-1)$, and the potential 
\begin{equation}\label{vrw}
V_\ell^-(r)=\left(1-\frac{2M}{r}\right)
\left[\frac{\ell(\ell+1)}{r^2}-\frac{6M}{r^3}\right].
\end{equation}
Because we are considering only axisymmetric perturbations
all components with $m\neq0$ vanish identically. We will subsequently
suppress the $m$ labels.

Moncrief showed that one can define a gauge-invariant function, that 
is invariant under infinitesimal coordinate transformations (gauge 
transformations), which is defined for {\em any} gauge via
\begin{equation}
\phi_{\ell}=\frac 1r\left( 1-\frac{2M}r\right) 
\left[ c_1^{(\ell)}+\frac 12\left( \partial
_rc_2^{(\ell)}-\frac 2rc_2^{(\ell)}\right) \right]  \label{Q}
\end{equation}
which satisfies the Regge-Wheeler equation above.  As the 
Regge-Wheeler equation is a wave equation, in order to evolve the 
function $\phi$ we must also provide its first time derivative, which is 
computed then directly through the definition of the extrinsic 
curvature of the perturbed Schwarzschild background:
\begin{eqnarray}  \label{Qpunto}
\partial _t\phi_{\ell}
&=&-\frac 2r\left( 1-\frac{2M}r\right) \left[ \sqrt{1-\frac{2M}r}%
K_{r\varphi }^{(\ell)\text{odd}}\right.\\
&&\left.+\partial _r\left( \sqrt{1-\frac{2M}%
r}K_{\theta \varphi }^{(\ell)\text{odd}}\right)-\frac 2r\sqrt{1-\frac{2M}r}%
K_{\theta \varphi }^{(\ell)\text{odd}} \right].\nonumber
\end{eqnarray}

This general prescription of linear Schwarzschild perturbations 
simplifies dramatically in the present case.  As discussed in 
Sec.~\ref{sec:initial} above, the three-metric contains only 
even-parity perturbations, and those appear {\em only} at second 
order.  Hence, for linearized treatment both the even- and odd-parity 
perturbation functions vanish in the initial data!  To first order the 
metric is described by the Schwarzschild background and perturbed 
initial data consists solely of odd-parity extrinsic curvature 
contributions.  Even-parity modes appear only at higher order, and are 
not considered in our comparisons here.

For the specific initial data given in the previous section we 
obtain 
\begin{eqnarray}
&&\phi_\ell\big|_{t=0}=0\\
&&\partial_t \phi_\ell\big|_{t=0}=-\frac{2}{r^3}\left( 1-\frac{2M}r\right)
\left\{-\{r\varphi\}_\ell\ q_G\right.\nonumber\\
&&\left.+\{\theta\varphi\}_\ell\left[
\partial^2_\eta q_G+\frac{(7M-3r)}{r\sqrt{1-\frac{2M}r}}\partial_\eta q_G
\right]\right\},
\end{eqnarray}
where for $n'=3,\ \ell=3$ and $n'=5,\ \ell=3,5$ the numerical coefficients
coming from the multipole decomposition of the extrinsic curvature are
\begin{eqnarray}
\{r\varphi\}_\ell&=&\frac{2^{(n'-3)/2}(4-\ell)}{(n'-2)(\ell-2)\ell(\ell+2)}\\
\{r\theta\}_\ell&=&\frac{2^{(n'-3)/2}(4-\ell)(\ell+1)}{(n'-2)(\ell-2)\ell}.\\
\end{eqnarray}
Although these 
initial data could be obtained numerically via an extraction process 
described in Paper I\cite{Allen97a}, it is not necessary to do so
in this case with a clear analytic linearization.

\begin{figure}
\epsfysize=3in \epsfbox{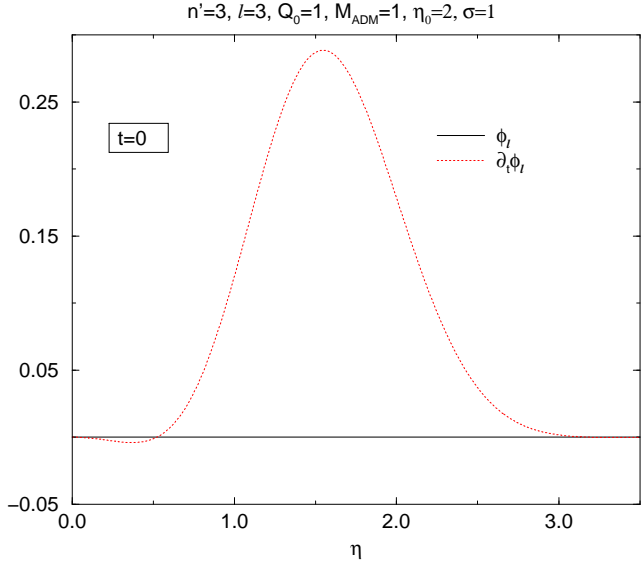} \caption{The initial data for the 
Moncrief variable.  For $n'=3$ the only linear content is the $\ell=3$ 
multipole.  The initial value of $\phi$ vanishes for linear odd-parity 
perturbations, since our choice of initial data only allows for second 
order even-parity perturbations of the initial three-metric.  Our 
choice of the initial extrinsic curvature generates an almost Gaussian 
$\partial_t\phi$ that sits near the maximum of the Regge--Wheeler 
potential.  }
\label{fig:carlos1}
\end{figure}

In Fig.\ \ref{fig:carlos1} we plot an example of these analytic
initial data.  We are now ready to evolve linearly these data with 
the Regge-Wheeler equation.

\subsubsection{Curvature Perturbations}
\label{sec:curv}
There is an independent formulation of the perturbation problem
derived from the Newman-Penrose formalism\cite{Teukolsky73} that is 
valid for perturbations of rotating black holes.
This formulation fully exploits the null structure of black holes 
to decouple the perturbation equations into a single wave equation 
that, in Boyer-Lindquist coordinates $(t,r,\theta,\varphi)$, can be 
written as:
\begin{eqnarray}
&&\ \ \Biggr\{\left[ a^2\sin^2\theta -\frac{(r^2+a^2)^2}\Delta \right]
\partial_{tt}-\frac{4Mar}\Delta \partial_{t\varphi } \nonumber \\
&&-2s\left[ (r+ia\cos
\theta )-\frac{M(r^2-a^2)}\Delta \right] \partial_t  \nonumber \\
&&\ \ +\,\Delta^{-s}\partial_r\left( \Delta^{s+1}\partial_r\right) +
\frac 1{\sin \theta }\partial_\theta \left( \sin \theta \partial_\theta
\right) \nonumber \\
&& +\left( \frac 1{\sin^2\theta }-\frac{a^2}\Delta \right) 
\partial_{\varphi \varphi }
+\,2s\left[ \frac{a(r-M)}\Delta +\frac{i\cos \theta }{\sin^2\theta }
\right] \partial_\varphi  \nonumber \\
&&-\left( s^2\cot^2\theta -s\right) \Biggr\}\psi
=4\pi \Sigma T \;,  \label{master}
\end{eqnarray}
where $M$ is the mass of the black hole, $a$ its angular momentum per unit
mass, $\Sigma  \equiv r^2+a^2\cos^2\theta$, and $\Delta 
\equiv  r^2-2Mr+a^2$.
The source term $T$ is built up from the energy-momentum
tensor\cite{Teukolsky73}.
Gravitational perturbations $s=\pm2$ are compactly 
described in terms of contractions of the Weyl tensor with a null 
tetrad, which components (also given in Ref.\ \cite{Teukolsky73}) 
conveniently chosen along the repeated principal null directions 
of the background spacetime (Kinnersley choice) 
\begin{equation}  \label{psi}
\psi (t,r,\theta ,\varphi )=\left\{ 
\begin{array}{ll}
\rho^{-4}\psi_4\equiv -\rho^{-4}C_{n\bar mn\bar m} & {\rm for}~~s=-2
\\ 
\psi_0\equiv -C_{lmlm} & {\rm for}~~s=+2~
\end{array}
\right. ,  \label{RH}
\end{equation}
where an overbar means complex conjugation and $\rho \equiv 
1/(r-ia\cos \theta )$. This field
represents either the outgoing radiative part of the perturbed Weyl tensor, 
($s=-2$), or the ingoing radiative part, ($s=+2$).


For the applications in this paper we will consider $s=-2$, since we 
will study emitted gravitational radiation and $a=0$, i.e. 
perturbations around a Schwarzschild black hole.  In general, for the 
rotating case, it is not possible to make a multipole decomposition 
of $\psi$ which is preserved in time.  So to keep generality we shall 
use Eqs.\ (3.1) and (3.2) of Ref.\ \cite{Campanelli98c} to build up our 
initial $\psi_4$ and $\partial_t\psi_4$
{\em not} decompose into $\ell$ multipoles.  The analytic 
expressions for the distorted black hole initial data sets considered 
in this paper, are:
\begin{eqnarray}\label{psinicial}
\psi_4\Big|_{t=0}&=&-\frac{i}{4}\frac{(1-2\frac{M}{r})}{r^4}
\cos{\theta}\sin^{n'-3}{\theta}\nonumber\\
&&\Big\{[(n'^2+n'-2)\sin^2{\theta}
-(n'^2-2n'-3)]q_G\nonumber\\
&&+ \sin^2{\theta}[4\sqrt{1-2\frac{M}{r}}\partial_\eta{q_G}-2\partial_\eta^2
{q_G}]\Big\}
\end{eqnarray}
\begin{eqnarray}
\partial_t\psi_4\Big|_{t=0}&=&\frac{i}{2}\frac{(1-2\frac{M}{r})}{r^7}
\cos{\theta}\sin^{n'-3}{\theta}\nonumber\\
&&\Big\{-\sqrt{1-2\frac{M}{r}}r^2\sin^{2}{\theta}
\partial_\eta^3{q_G}\nonumber\\
&&+(5r-12M)r\sin^{2}{\theta}
\partial_\eta^2{q_G}\nonumber\\
&&+r\sqrt{1-2\frac{M}{r}}[(21M-8r+rn'^2+rn')\sin^{2}{\theta}\nonumber\\
&&+(3+2n'-n'^2)r]\partial_\eta{q_G}+[(n'^2+n'-2)\sin^{2}\theta\nonumber\\
&&-(n'^2-2n'-3)](5M-2r)r\ q_G\Big\}.
\end{eqnarray}

\begin{figure}
\epsfysize=3in \epsfbox{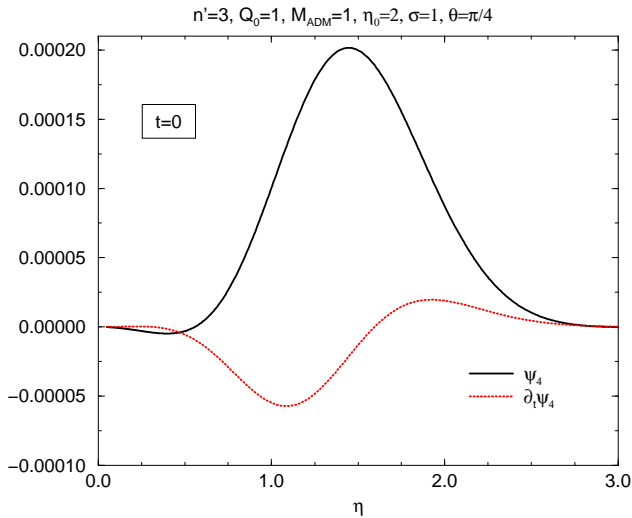}
\caption{The initial data for the Weyl scalar $\psi_4$ and its time derivative
as given in Eq.\ (\protect\ref{psinicial}).
The general form of $\psi_4$ resembles that of the Gaussian-like
$\partial_t\phi_\ell$. Note that $\partial_t\psi_4$ is nonvanishing
in contrast to the initial $\phi_\ell=0$
}
\label{fig:carlos2}
\end{figure}

As expected for pure odd-parity perturbations, only the imaginary part of
$\psi_4$ is nonvanishing.  Also note that unlike $\phi_\ell=0$,
$\partial_t\psi_4$ does not vanish initially. This is because $\psi_4$ and
its time derivative
depend on both, the 3-geometry and the extrinsic curvature
while the Moncrief function only depends on the perturbed 3-geometry and
its time derivative only on the extrinsic curvature.
We plot the initial data in Fig.\ \ref{fig:carlos2}.
We evolve of these initial data via Teukolsky equation(\ref{master}) using
the numerical method is described
in Ref.\ \cite{Krivan97a}. 

Since, after all, we are computing perturbations on a Schwarzschild ($a=0$)
background there must be a way to relate the metric and curvature
approaches. In fact the relations between $\psi_4$ and Moncrief even- and
odd-parity waveforms in the time domain have been found in 
Ref.\ \cite{Campanelli98a} and tested for the even-parity case in 
Ref.\ \cite{Campanelli98b}\footnote{Note that this relation among waveforms
is only valid at first perturbative order. When nonlinearities are
included the two approaches may give widely different results
\protect\cite{Lousto99a}.}

Here we can perform the same kind of cross check
for the odd parity modes.
From the equations in section II.B of Ref.\ \cite{Campanelli98a} or (2.9) in 
Ref.\ \cite{Campanelli98b} we obtain a relation that holds at {\it all times}
\begin{eqnarray}\label{chandra}
&&{\partial_t\psi}_4(t,r,\theta,\varphi)=
-\frac{i}{8r^2}\sum_{\ell}\sqrt{\frac{(\ell+2)!}{(\ell-2)!}}
\ {}_{-2}Y_\ell(\theta,\varphi)\times\nonumber\\
&&\Bigg\{2r\partial_{r^*}\Big[\partial_t\phi_{\ell}(t,r)-
\partial_{r^*}\phi_{\ell}(t,r)\Big]\\
&&+2(1-\frac{3M}{r})
\Big[\partial_t\phi_{\ell}(t,r)-
\partial_{r^*}\phi_{\ell}(t,r)\Big]+rV_\ell^-\phi_{\ell}(t,r)\Bigg\}\nonumber
\end{eqnarray}
and that we can integrate to give us $\psi_4$ from $\phi_\ell$ evolved with
the Regge-Wheeler (\ref{rtrw}) equation instead of evolving $\psi_4$ directly
with the Teukolsky equation (\ref{master}).

\subsection{Axisymmetric Nonlinear Evolutions}
\label{sec:2Dnonlinear}

The  2D fully nonlinear evolutions have been performed with a code,
{\sl Magor}, designed to evolve axisymmetric, rotating, highly 
distorted black holes, as described in 
Ref.~\cite{Brandt94a,Brandt94b,Brandt94c}. {\sl Magor} has also 
been modified to include matter flows accreting onto black 
holes~\cite{Brandt98}, but here we consider only the 
vacuum case.    

In a nutshell, this nonlinear code solves the complete set of
Einstein equations, in axisymmetry, with maximal slicing, for a 
rotating black hole.  The code is written in a spherical-polar coordinate 
system, with the rescaled radial coordinate $\eta$ that vanishes on the 
black hole throat.  An isometry operator is used to provide boundary 
conditions on the throat of the black hole.  All three components of a 
shift vector are employed to keep all off diagonal components of the 
metric zero, except for the $g_{\theta \varphi}$ component, which carries 
information about the odd-parity polarization of the radiation.  For 
complete details of the nonlinear code, please see 
Refs.~\cite{Brandt94a,Brandt94b,Brandt94c,Brandt98}.

The initial data described in Sec.~\ref{sec:initial} above are provided 
through a fully nonlinear, numerical solution to the Hamiltonian constraint.
  The code is able to evolve such data sets for time scales of 
roughly $t\leq 10^2M$, and study such physics as horizons and gravitational 
wave emission.

Consistently with the two different perturbations approaches there 
are two methods we use to extract information about the 
gravitational waves emitted during the fully nonlinear simulation: 
metric based gauge-invariant waveform extraction and direct evaluation 
of the curvature based Newman-Penrose quantities, such as $\psi_{4}$.

The first method has been developed and refined over the years 
~\cite{Abrahams88b,Seidel92c,Moncrief74,Moncrief74b,Zerilli70,Regge57}
to compute waveforms from the numerically evolved metric. 
Surface integrals of various metric quantities are combined to build up the 
perturbatively gauge invariant odd-parity Moncrief functions. These can then 
be compared directly with the perturbative results.

A second method for wave extraction is provided by the calculation 
of the Weyl scalar $\psi_4$ which is coordinate invariant but depends on a
choice of tetrad basis. For our numerical extractions we follow the method
proposed in Ref.\ \cite{Gunnarsen95}. 
To define the tetrad of their form numerically  we thus align the real
vector (which can be 
thought of as providing the spatial components of $\l^\mu$ and 
$l^\mu$) with the radial direction.  The complex vectors $m^\mu$ and
$\bar{m}^\mu$ point within the spherical 2-surface.  At each step,
a Gram-Schmidt procedure is used to ensure that the triad remains
ortho-normal.   The tetrad assumed by this method is not directly 
consistent with the one assumed in the perturbative calculation, but for
the $a=0$ it can be made consistent by a type III (boost) null rotation
which fixes the relative normalization of the two real-valued vectors.
We have found that the transformation $n_P\to A^{-1}l_O$ and $l_P\to Al_O$
where $A=\sqrt{2/(1-2M/r)}$ fixes the normalization appropriately.  For the
general $a\neq 0$ case this would be insufficient, and we would instead use
the more general method proposed in Ref.\ \cite{Campanelli98c}.


\subsection{Full 3D Evolutions with Cactus}
\label{sec:3Dnonlinear}

The last of our approaches for evolving these distorted black hole 
data sets utilizes full 3D nonlinear numerical relativity, and is 
based on {\sl Cactus}.  More general than a numerical code, the {\sl Cactus} 
Computational Toolkit is actually a general parallel framework for 
numerical relativity (and other sets of PDE's), that allows users from 
various simulation communities to gain high performance parallelism on 
many platforms, access a variety of computational science tools, and 
to share modules of different evolution methods, initial data, 
analysis routines, etc.  For the relativity community, an extensive 
suite of numerical relativity modules (or thorns in the language of 
{\sl Cactus}) is available\footnote{
{\sl Cactus} is well documented and can be downloaded freely 
from a web server at http://www.cactuscode.org.  For more information 
on {\sl Cactus}, its use in numerical relativity and other fields, please 
see the web pages.}, including black hole and other initial data, 
slicing routines, horizon finders, radiation indicators, evolution 
modules, etc.

For this paper, {\sl Cactus} was used to assemble a set of 3D initial data, 
evolution modules, and analysis routines needed for the comparisons 
with {\sl Magor} and the two perturbative approaches described above.  All 
operations have been carried out in 3D Cartesian coordinates, from 
initial data to evolution to waveform extraction.  The initial data 
are computed as in the Magor code, in a polar-spherical type 
coordinate system, and interpolated onto the Cartesian coordinate 
system as described in Paper I. The evolutions are carried out with a 
formulation of Einstein's equations based on the conformal, trace-free 
approach developed originally by Shibata and Nakamura~\cite{Shibata95} 
and Baumgarte and Shapiro~\cite{Baumgarte99}, and further tested and 
developed by Alcubierre, et.  al., as described 
in~\cite{Alcubierre99e,Alcubierre99d}.  Due to certain symmetries in 
these initial data sets, the evolutions can be carried out in an 
octant, in Cartesian coordinates.  However, we have chosen in this 
case to use the full 3D Cartesian grid, as enough memory is now 
available to run sufficiently large scale evolutions that cover the 
entire spacetime domain of interest, as would also be necessary when 
considering the general black hole inspiral problem.  For comparison with
the 2D code we extract waveforms via the same gauge-invariant Moncrief
approach.  In this case, 
surface integrals are carried out on the Cartesian based system by 
coordinate transformations and interpolation onto a coordinate 
2--sphere, as described in Ref.~\cite{Allen98a}.  

Further details of the individual simulation parameters are provided 
as needed when discussing the results below.

\begin{figure}
\epsfysize=2.8in \epsfbox{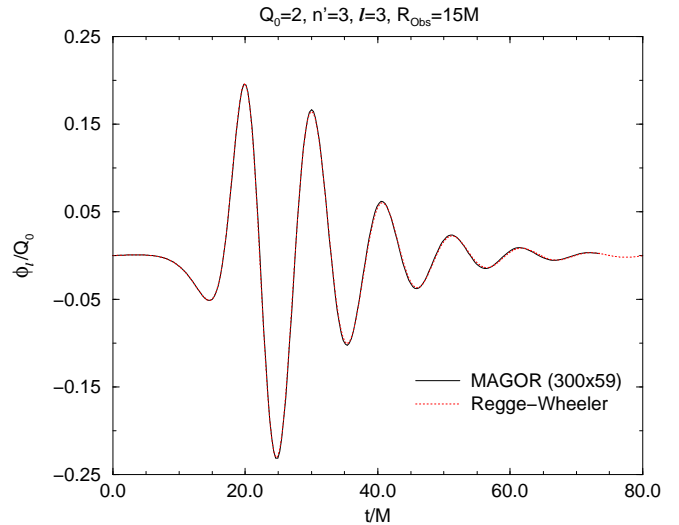}
\caption{The $\ell=3$, odd-parity Moncrief waveform, extracted 
from the fully nonlinear 2D  evolution code (solid line) and compared 
to the fully linear evolution of the same data, obtained by the 
Regge-Wheeler equation, for the low amplitude case 
$(Q_{0}=2,n'=3,\eta_{0}=2,\sigma=1)$.  The excellent agreement shows 
that both approaches are accurate, and that the black hole physics is 
operating in the linear regime.
}
\label{fig:ed1}
\end{figure}

\section{Results}

Here we compare the results of evolving the odd-parity distorted black holes
by the four techniques described above.  We consider two classes of
distortions ($n'=3$ and $n'=5$) with different angular distributions, and
various amplitudes to include cases of linear and distinctly nonlinear
dynamics.  For the $n'=3$ case the distortion is (linearly) pure $\ell=3$,
while the $n'=5$ case encodes a mix of $\ell=3$ and $\ell=5$ distortions in
the initial data.

\subsection{Comparison of Nonlinear Evolutions with Regge-Wheeler 
Theory}
In this subsection we compare the 2D nonlinear ({\em Magor}) evolutions
with the results of the Regge-Wheeler perturbative approach. 
We first consider the nonlinear evolution of a family of data sets with 
parameters  $(Q_{0},n'=3,\eta_{0}=2,\sigma=1)$.

For low amplitude cases with $Q_0<8$, we are in the linear regime and even 
the nonlinear evolutions exhibit strongly linear dynamics.  In 
Fig.~\ref{fig:ed1} we show $\ell=3$ waveform results 
obtained from the 2D nonlinear code, for the case $Q_{0}=2$, and 
compare with Regge-Wheeler evolutions of the $(\phi,\partial_{t}\phi)$ 
system.  The agreement is so close that the curves cannot be distinguished 
in the plot.  The perturbative-numerical agreement is equally good with
the other linear waveforms at low amplitude so we will leave the
perturbative results out of the plots and focus on the transition to
nonlinear dynamics.

\begin{figure}
\epsfysize=2.8in \epsfbox{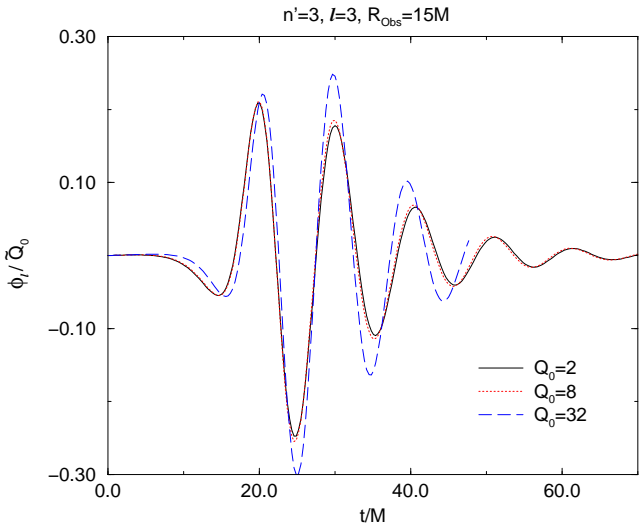}
\caption{The $\ell=3$, odd-parity Moncrief waveform, extracted 
from the fully nonlinear 2D  evolution code, for a series of 
evolutions with initial data parameters 
$(Q_{0},n'=3,\eta_{0}=2,\sigma=1)$.  The waveforms are extracted at
isotropic coordinates $\bar{r}=15M$.
For $n'=3$ we only have $\ell=3$ linear contributions. We normalized
waveforms by the amplitude $\tilde Q_0$ in order to study the linear and
nonlinear regimes.
It is observed that for $Q_0\leq8$ the linear regime is maintained, while for
$Q_0=32$ nonlinearities are well noticeable. The effect of nonlinear
contributions increases the scaled amplitude of the waveform and increases
its frequency. This indicates that the final ringing black hole is
significantly smaller than the initial mass of the system.
}
\label{fig:john1}
\end{figure}

In Fig.~\ref{fig:john1} we show the $\ell=3$ gauge-invariant 
Moncrief waveforms 
for a sequence of such evolutions of increasing amplitude $Q_{0}$.  
The waveforms have all been normalized by the amplitude factor 
$\tilde Q_{0}=Q_0/M^2$
to accentuate nonlinear effects.  If the system is in the 
linear regime, the normalized waveforms will all line up, as is 
clearly the case in the regime $Q_{0}\le8$.  For the large amplitude 
case $Q_{0}=32$, the normalized waveform is much larger, indicating 
that here we are well into the nonlinear regime.

\begin{figure}
\epsfysize=2.8in \epsfbox{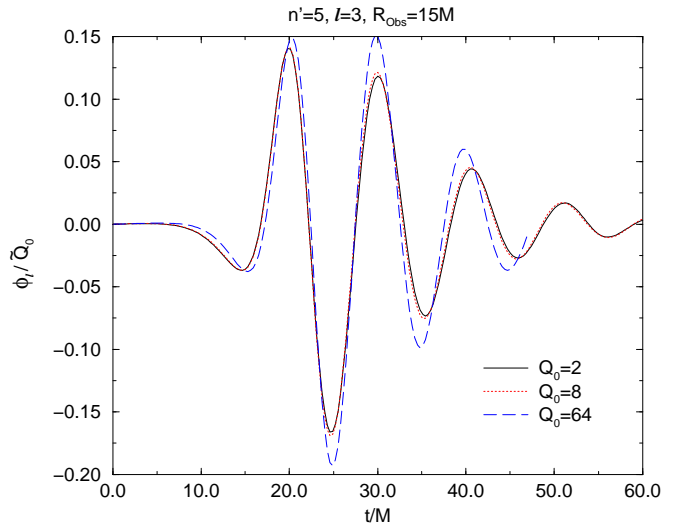}
\caption{We show the $\ell=3$ normalized odd-parity Moncrief waveform for
$(Q_{0},n'=5,\eta_{0}=2,\sigma=1)$ initial data, extracted from the 
fully nonlinear 2D code for a variety of amplitudes $Q_{0}$.
Again, the system is clearly linear for $Q_0\leq8$ and nonlinearities cause 
an increase in amplitude and 
frequency of the wave.
}
\label{fig:john2}
\end{figure}

\begin{figure}
\epsfysize=2.8in \epsfbox{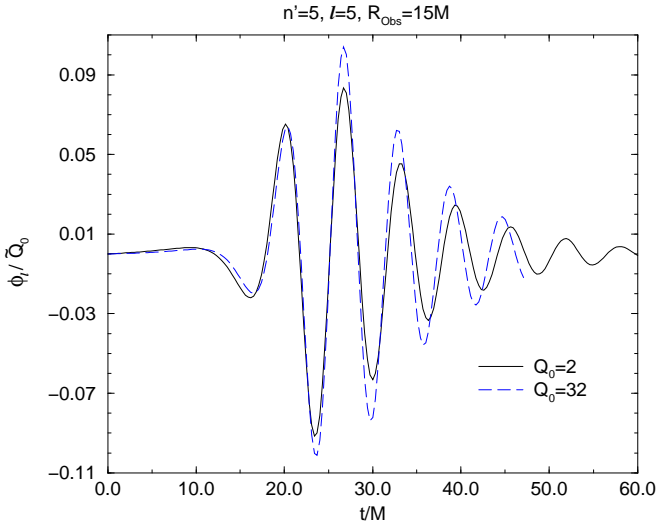}
\caption{We show the $\ell=5$ normalized odd-parity Moncrief waveform for
$(Q_{0},n'=5,\eta_{0}=2,\sigma=1)$ initial data, extracted from the 
fully nonlinear 2D code for a variety of amplitudes $Q_{0}$.
The regime is clearly linear for $Q_0=2$ and nonlinear components appear for
$Q_0=32$. The increase in frequency and amplitude of the wave are seen
here also.
}
\label{fig:john3}
\end{figure}

We now consider the transition to nonlinear dynamics in our second 
family of data sets, given by parameters 
$(Q_{0},n'=5,\eta_{0}=2,\sigma=1)$.  These data sets have a linear 
admixture of both $\ell=3$ and $\ell=5$ perturbations, and should 
contain waveforms of both types.  In Fig.~\ref{fig:john2}, we show 
the results of the fully nonlinear evolution with the {\sl Magor} code in 
maximal slicing, extracting the $\ell=3$ gauge-invariant Moncrief 
waveform for various amplitudes.  The waveforms are again normalized 
by the amplitude factor $\tilde Q_{0}$.  In this case the nonlinearity
is somewhat weaker at $Q_{0}=32$ so we have included the $Q_{0}=64$ curve
in the figure.  For the $\ell=5$ case, shown in Fig.~\ref{fig:john3}, the
higher frequency of the quasi-normal ringing makes it easier to appreciate
the nonlinearities at $Q_{0}=32$.  The plots indicate that again the
dynamics are quite linear below $Q_{0}=8$

\begin{figure}
\epsfysize=2.8in \epsfbox{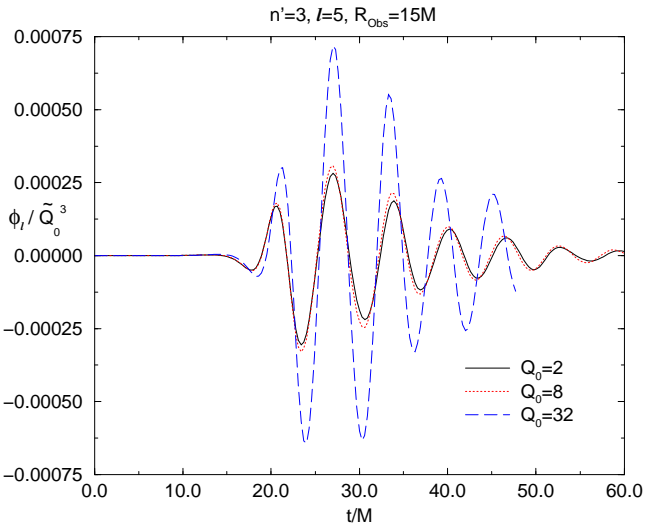}
\caption{The Moncrief waveform for a purely nonlinear mode.
For $n'=3$, the $\ell=5$ multipole is generated by cubic products of the
odd-parity wave (squares generate even-parity ones). Accordingly we normalized
waveforms by $Q_0^3$. Still higher nonlinearities switch on for $Q_0=32$ and
show the generic increase in the frequency of the wave.
}
\label{fig:john4}
\end{figure}

The waveforms we have shown so far are the only ones predicted to linear 
order in perturbation theory.  We would need to apply higher order 
perturbation theory to predict waveforms for the even-parity or higher-$\ell$
 odd parity modes. Nevertheless general considerations from the perturbative 
point of view do provide some expectations on the scaling of the other 
nonlinear waveform modes within the families considered here.  We return to 
the $n'=3$ family for an example.  
The leading contribution to the 
$n'=3$, $\ell=5$ odd-parity waveform comes from the cubic coupling 
of the first order $\ell=3$ odd-parity mode discussed above
(including the coupling of the $\ell=3$ 
odd-parity mode with the second order 
even-parity $\ell=2$ mode expected via the source term contribution to the 
solution of the Hamiltonian constraint in the initial data. 
Thus, this
wave component should appear at the third perturbative order.  We verify this 
expectation by plotting the numerical results for the  $\ell=5$ odd-parity 
waveforms scaled this time by ${\tilde Q_0}^3$ in Fig~\ref{fig:john4}.  
Although the magnitudes of these waveforms are far smaller than those of 
the $\ell=3$ mode we again see very nice agreement, below $Q_0=8$, with 
the perturbative expectation, that the waveforms should superpose.

Let us now take another look at Figs. \ref{fig:john1}-\ref{fig:john3} to 
consider what is happening as we move into the nonlinear regime where 
the waveforms no longer superpose.  In all the graphs we see the same 
general features arising as begin to drive the system into the nonlinear 
regime.  These are higher frequency ringing, and larger amplitudes for the 
later parts of the waveform.  At $Q_0=32$ the $n'=3$ case shows a roughly 10\% 
increase in frequency compared to 5\% for the $n'=5$ case.  Since the final
state of the system will be a black hole, we expect quasi-normal ringing in
the late-time behavior of the system regardless of the size of the initial
perturbations.  This is indeed what we see in the waveforms, except that the
ringing is at a higher frequency (relative to the initial mass of the system)
than we expected.  This indicates that the mass of the final ringing black
hole has less mass than the ADM mass of the initial data.  The perturbations
have grown large enough to generate radiation amounting to a noticeable
fraction of the total ADM mass leaving behind a slightly smaller black hole.
The smaller mass of the final black hole is also consistent with a larger
amplitudes, since the scaled perturbation $\tilde Q_0=Q_0/M^2$ is larger
relative to a smaller mass black hole.  The arrival time of the wave pulse
is not strongly affected by the change in mass because the time and wave
extraction points are both scaled against the initial mass.

\subsection{Comparison of Nonlinear Evolutions with Teukolsky 
Theory}

We now turn to the curvature based Teukolsky approach to perturbative 
evolution for black hole spacetimes.  As motivated above, 
this is a much more powerful 
approach that will enable perturbative evolutions of both 
polarizations of the gravitational wave, and evolutions of distorted black 
holes with angular momentum, and without the need for multipolar 
expansions.  The key difference  for analysis within this (Teukolsky) 
formalism is that , in the general case, we no longer have the benefit 
of a time-independent separation into multipoles.   In this first step 
towards the transition between the 
metric perturbation (Moncrief) approach and the curvature based 
(Teukolsky) approach, we consider the same data sets studied above, 
which have linear perturbations only for odd-parity, nonrotating 
black holes.  We will consider more general systems in future papers.

\begin{figure}
\epsfysize=3in \epsfbox{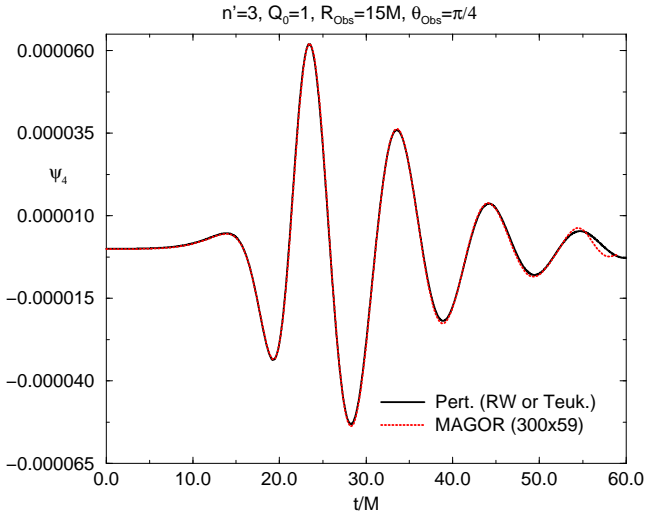}
\caption{The Weyl scalar $\psi_4$ as seen by an observer
located at isotropic coordinates $\bar{r}=15M$ and $\theta=\pi/4$.
For $n'=3$ we have a pure $\ell=3$ linear contribution although no
multipole decomposition was made of $\psi_4$. We compare here the results
of integrating full (axially symmetric) Einstein equations numerically
with the program {\sl Magor} using a resolution of 300 radial and 59 angular
zones with the linear evolution of initial data via the Teukolsky equation
or the Regge-Wheeler equation and then use the transformation equations
(\protect\ref{chandra}) to
build up $\psi_4$ [only possible in the nonrotating case].
}
\label{fig:carlos3}
\end{figure}

We now consider evolutions of the distorted black hole data set 
$(Q_{0} =1,n'=3,\eta_{0}=2,\sigma=1)$.  
The perturbative initial data for $\psi_{4}$ and 
$\partial_{t}\psi_{4}$, needed for use in the Teukolsky evolution, 
have been obtained as described in Sec.~\ref{sec:curv} above.  These 
data are then evolved and recorded at the same coordinate location as 
before ($r=15M$) for comparison with the previous results.  In Fig.\ 
\ref{fig:carlos3} we show the results of the full 2D nonlinear evolution, 
obtained with {\sl Magor} in maximal slicing, the Teukolsky evolution, and 
the Regge-Wheeler evolution obtained previously. The solid line shows 
the results of $\psi_{4}$ obtained with the Teukolsky equation, 
observed at a constant angular location $\theta=\pi/4$, and the 
dotted line shows the result of the {\sl Magor} evolution at the same 
location, with $\psi_{4}$ extracted from the full nonlinear 
simulation as described in Sec.\ II.C.2 above.  The results agree 
extremely well except at very late times, when the nonlinear results 
are affected slightly by coarse resolution in the outer regions of the 
numerical grid.  We also verify here that the  results of the Regge-Wheeler 
evolution, transformed to provide the same function $\psi_{4}$ 
according to Eq.~(\ref{chandra}) above, agree with the Teukolsky evolution.
The results of the two perturbative approaches are indistinguishable in the
graph.

\begin{figure}
\epsfysize=3in \epsfbox{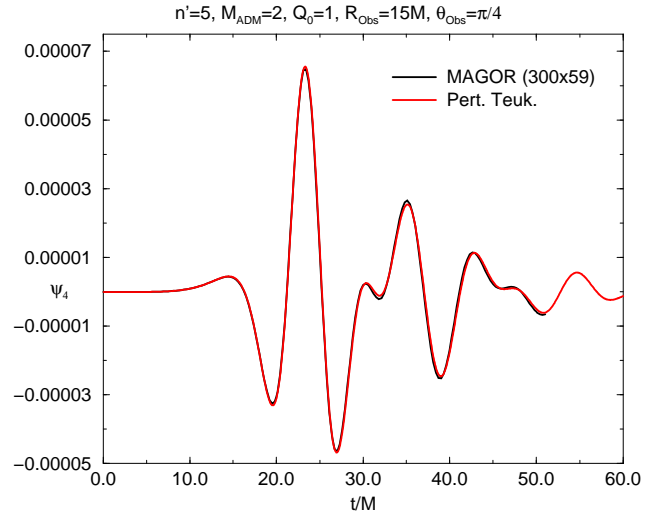}
\caption{The Weyl scalar $\psi_4$ for the $n'=5$ initial data. We observe the
beating of the $\ell=3$ and $\ell=5$ components since we are not making any
multipole decomposition. 
}
\label{fig:carlos4}
\end{figure}

We now examine the other family of distorted black holes with the choice
of angular parameter $n'=5$.  The initial data were 
obtained as before, and evolved with the nonlinear {\sl Magor} code and the 
Teukolsky code.  The results are shown in Fig.\ \ref{fig:carlos4}, 
where we see excellent agreement between the two plots.  But notice that 
the waveform does {\em not} show the clear quasi-normal mode 
appearance that one is accustomed to in such plots.  This is because 
this data set has a roughly equal admixture of both $\ell=3$ and 
$\ell=5$ components of radiation, and the curvature based $\psi_{4}$ 
approach is not decomposed into separate multipoles.  This 
waveform shows a clear beat of the two $\ell=3$ and $\ell=5$ components.

\subsection{Comparison of nonlinear 2D and full 3D codes}

Having successfully tested the 2D fully nonlinear code {\sl Magor} for
odd-parity distortions against perturbative evolutions, we can now test
the 3D code {\sl Cactus} against the 2D one. In {\sl Cactus}, using the same
procedures as in Sec.~\ref{sec:2Dnonlinear}, the initial data are
evolved in the full (no octant) 3D mode, with a second order
convergence algorithm, maximal slicing, and static boundary conditions.
Note we perform the conformal-traceless 
scheme ~\cite{Alcubierre99e,Alcubierre99d} for this evolution.
The first observation is that we have to solve the initial value
problem taking into account all nonlinearities, even if we are in the
linear regime $(Q_0<10)$, since small violations of the Hamiltonian
constraint contaminate the outgoing waveforms.

\begin{figure}
\epsfysize=2.8in \epsfbox{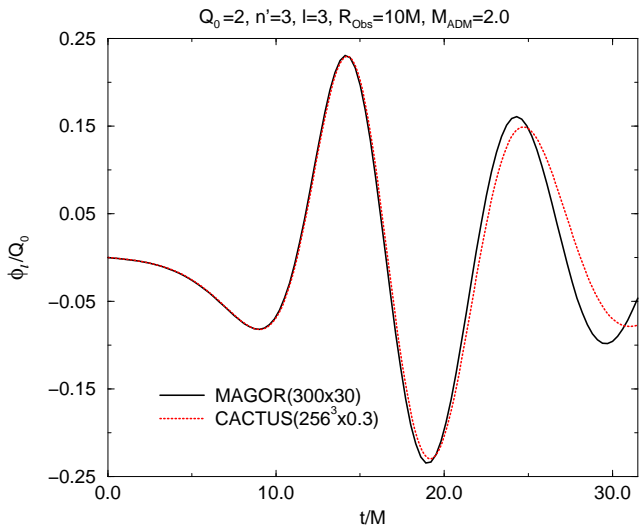}
\caption{
The $\ell=3$, odd-parity Moncrief waveform, extracted 
from the fully nonlinear 3D evolution code {\sl Cactus} (dotted line).
The spatial grid consists of $256^3$ points with a separation of $0.15M$.
The ADM mass of the black hole is $M=2.0$ and
$(n'=3,Q_{0}=2,\eta_{0}=2,\sigma=1)$.
For comparison we also plot the results of evolving the same initial
data with the fully nonlinear 2D evolution code {\sl Magor} (solid line), which
in turn has been tested against perturbation theory as shown in
Fig.\ \protect{\ref{fig:ed1}}.
Very good agreement is reached with a relatively low resolution
of the 3D code.
}
\label{fig:ryoji2}
\end{figure}

\begin{figure}
\epsfysize=2.8in \epsfbox{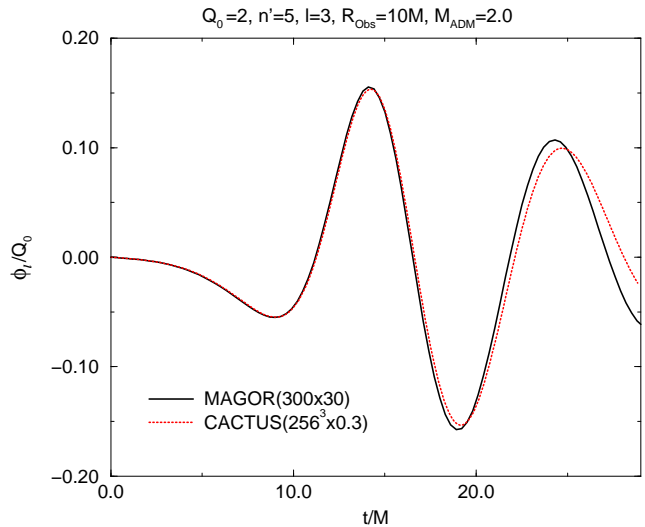}
\caption{The $\ell=3$, odd-parity Moncrief waveform produced by
the 3D code {\sl Cactus} (dotted line). It corresponds to initial data with
parameters $(n'=5,Q_{0}=2,\eta_{0}=2,\sigma=1)$.
It also show very good agreement with the 2D results (solid line)
which had been checked against perturbation theory as displayed in
Fig.\ \protect{\ref{fig:john2}}.
}
\label{fig:ryoji3}
\end{figure}

\begin{figure}
\epsfysize=2.8in \epsfbox{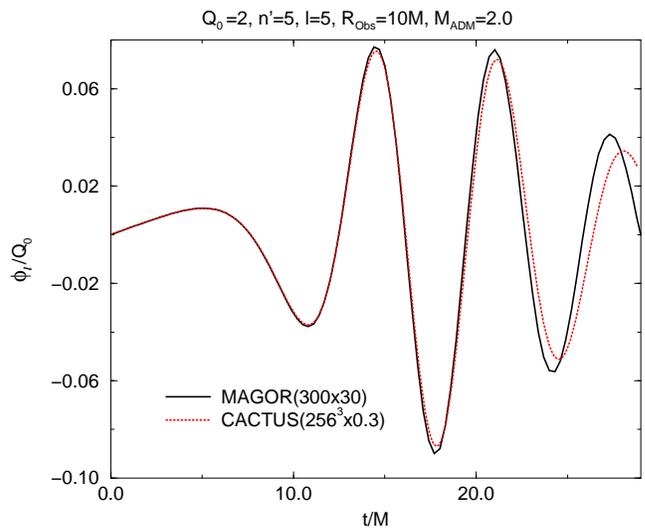}
\caption{
The $\ell=5$, odd-parity Moncrief waveform produced by
the 3D code {\sl Cactus} (dotted line) with initial data having
$(n'=5,Q_{0}=2,\eta_{0}=2,\sigma=1)$.
Comparison with the 2D results (solid line) show a very good agreement.
See Fig.\ \protect{\ref{fig:john3}} for the agreement between the 2D run
and perturbation results.
}
\label{fig:ryoji4}
\end{figure}

The runs presented in
Figs.\ \ref{fig:ryoji2}, \ref{fig:ryoji3}, and \ref{fig:ryoji4} show
very nice agreement with the 2D code (hence also with perturbation theory).
Note that the spatial resolution $(\Delta x^j=0.15M=0.3)$ is not high. Here
we show waveforms for $t/M\leq30$. The runs do {\it not} crash afterwards, but
 become less accurate due to the low resolution and boundary effects,
and even later to collapse of the lapse.

\begin{figure}
\epsfysize=2.8in \epsfbox{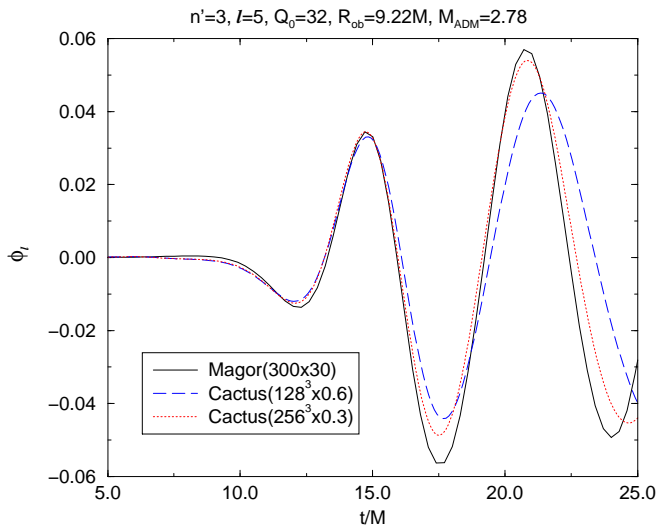}
\caption{
The $\ell=5$, odd-parity Moncrief waveform produced by
the 3D code {\sl Cactus} (dotted line) with initial data having
$(n'=3,Q_{0}=32,\eta_{0}=2,\sigma=1)$ and $M_{ADM}=2.777$. This is
a purely nonlinear mode, its
leading term being cubic in the amplitude $Q_0$.
Comparison with the 2D results (solid line) show a good rate of convergence
with the grid spacing (from $\Delta x=0.6M/2.777$ to $\Delta x=0.3M/2.777$.
See Fig.\ \protect{\ref{fig:john4}} for the purely 2D runs.
}
\label{fig:ryoji1}
\end{figure}

The $\ell-$modes shown in Figs.\ \ref{fig:ryoji2}-\ref{fig:ryoji4} are
essentially dominated (for $Q_0=2)$ by the linear initial distortion of
the black hole. Those are the modes that we can compare with first order
perturbation theory. Since we have two nonlinear codes we can now compare
their predictions for modes dominated by nonlinear effects. That is the case
of the odd mode $\ell=5$ when the initial data parameter is $n'=3$. This mode
has a linear contribution only for $\ell=3$. For $\ell=5$ is easy to see that
to generate an odd mode we need at least cubic contributions. Thus this mode
will scale as $Q_0^3$. To be able to verify the agreement between the 2D and
3D codes we amplified this mode taking $Q_0=32$ and checked the (almost)
quadratic convergence of {\sl Cactus} to the correct results as shown in
Fig.\ \ref{fig:ryoji1}.

\section{Discussion}

We have completed a series of comparisons covering four different 
approaches for two classes of odd-parity distortions of Schwarzschild 
black holes.  This includes 2D and 3D nonlinear evolutions and for the
first time, in both cases, a comparison of the odd-parity 
Regge-Wheeler-Moncrief formulation as well as the Teukolsky approach 
with numerical results.  In all cases we find excellent agreement among 
the different approaches.  We emphasize that these matchings have been
achieved without the aid of any parameters and thereby stand as a strong 
verification of these techniques.  

Although the distorted black hole initial data configurations we consider
here are not necessarily astrophysically relevant, our analysis provides
an example of the usefulness of perturbation theory as an interpretive
tool for understanding the dynamics produced in fully nonlinear evolutions.
In order to distinguish the cases of linear and nonlinear dynamics we
simply show the output of the full nonlinear code, but we scale
it by the factor $Q_0/M^2$ so that, if the system is responding
linearly to $Q_0$ all the waveforms will lie exactly on top of
one another.  Using this procedure we are able to recognize the emergence
of nonlinear dynamics.
Considering the mixing of perturbative modes also enables us to understand
the results of one case which displays strictly nonlinear behavior, the
$\ell=5$ waveform of the initial data with $n^\prime=3$
(see Figs.\ \ref{fig:john4} and \ref{fig:ryoji1}).
This wave strictly vanishes to linear order in $\tilde Q_0$ and
scales at lower amplitudes like $\tilde Q_0^3$.  The perspective of
perturbation theory allows us to create a full picture, identifying and
explaining aspects of the nonlinear dynamics even when the perturbations
are beyond the linear regime.  In this case we find that linearized
dynamics provide a very good approximation of the systems' behavior until
the radiation constitutes a significant portion of the initial mass,
producing a smaller final black hole and, for example, higher quasi-normal
ringing frequencies.

Although we restrict to the case where the black hole system does not 
have net angular momentum, the approach we develop in this 
paper is completely general, and can easily be extended 
to the case of distorted black holes with nonvanishing 
angular momentum.  For this reason, we developed a 
procedure for using the Teukolsky equation to evolve 
the perturbations on a black hole background,  
handling both the even- and odd-parity perturbations 
simultaneously, and  providing the capability to 
deal with perturbations evolving on a Kerr background.

In future work we expect to move in two directions:  
(a)  We will apply the techniques developed here to the case of 
distorted, rotating black holes, to study nonlinear effects in the
radiation of energy and angular momentum as well as to
further develop the Teukolsky perturbative evolution paradigm for
application to coalescing black hole initial data in a close-limit
approximation.  (b)  We will use these 
techniques to evolve black hole systems, either from numerically 
generated initial data, or from partially evolved datasets that have 
reached a stage where they can be treated via perturbation theory.

\acknowledgements 
We would like to thank our colleagues at AEI and Penn State, especially
Gabrielle Allen.
This work was supported by AEI, and by NSF grant PHY-0800973.
The nonlinear computations have been performed on a 8 Gbyte SGI Origin 2000
with 32 processors at AEI and a 98 Gbyte Cray T3E with 512 processors at
MPI-Garching.

\bibliographystyle{prsty}
\bibliography{references}
\thebibliography{odd2}

\end{document}